\documentclass[amsmath,amssymb,twocolumn,nofootinbib]{revtex4-1}
\usepackage[dvipsnames]{xcolor}
\usepackage{graphicx,epsfig,hyperref}
\usepackage{multirow}
\hypersetup{unicode=true,colorlinks=true,linkcolor=NavyBlue,citecolor=NavyBlue,urlcolor=black}
\usepackage[normalem]{ulem}
\usepackage{tikz}  
\usetikzlibrary{arrows,positioning,shapes,backgrounds,fit}
\begin{document}
\title{Looking for Carroll particles in two time spacetime}
\author{Alexander Kamenshchik}
\email{kamenshchik@bo.infn.it}
\affiliation{Dipartimento di Fisica e Astronomia, Universit\`{a} di Bologna,
via Irnerio 46, 40126 Bologna, Italy \\ I.N.F.N., Sezione di Bologna, I.S. FLAG, viale Berti Pichat 6/2, 40127 Bologna, Italy}
\author{Federica Muscolino}
\email{federica.muscolino@gmail.com}
\affiliation{Department of Science and High Technology, Universit\`a dell'Insubria, Via Valleggio 11, IT-22100 Como,Italy\\
I.N.F.N,  Sezione di Milano, via Celoria 16, IT-20133 Milano, Italy}
\begin{abstract}
We make an attempt to describe  Carroll particles with a non-vanishing value of energy (i.e. the Carroll particles which always stay in rest) in the framework of
two time physics, developed in the series of papers by I. Bars and his co-authors. In the spacetime with one additional time dimension and one additional space dimension 
one can localize the symmetry which exists between generalized coordinate and their conjugate momenta. Such a localization implies the introduction  of the gauge fields, which in turn implies the appearance of some first-class constraints. Choosing different gauge-fixing conditions and solving the constraints one obtain different time parameters, Hamiltonians, and generally, physical systems in the standard one time spacetime. In this way such systems as non-relativistic particle, relativistic particles, hydrogen atoms and harmonic oscillators were described as dual systems in the framework of the two time physics. Here, we find a set of gauge fixing conditions which provides as with such a parametrization of the phase space variables in the two time world which gives the description of Carroll particle in the one time world. Besides, we construct  the quantum theory of such a particle using an unexpected  correspondence between our parametrization and that obtained by Bars for the hydrogen atom in 1999.  
\end{abstract}
\maketitle

\section{Introduction}
The theories with extra spatial dimensions have become quite traditional since the times when they were put forward in the famous works by Kaluza and Klein. The theories with more than one time dimensions look much less intuitive and plausible.
However, an impressive series of papers devoted to the so called two time or 2T physics was produced by I. Bars and his co-authors \cite{Bars0,Bars01,Bars02,Bars,Bars1,Bars2,Bars3,Bars4,Bars5,Bars6,Bars7,Bars8,Bars9,Bars10,Bars94,Bars95,Bars96,
Bars97,Bars98,Bars99,Bars100,Bars-book}. 

The 2T-physics was born in the area of the cutting-edge research of the modern theoretical physics including supersymmetry, strings, branes, unified theories and all that \cite{Bars0,Bars01,Bars02,Bars,Bars1,Bars2}. A little bit later it was understood, 
that the classical and quantum physics of very simple systems such as non-relativistic particle, massive and massless relativistic 
particles, harmonic oscillator, hydrogen-like atoms can be described in the framework of 2T-physics from a unifying point of view
\cite{Bars3,Bars4,Bars5,Bars6}. As was described in the book \cite{Bars-book} different one time physical systems arise as some ``shadows'' in the Plato's cave, which is nothing but the world with one additional temporal dimension and one additional spatial 
dimension. Thus, some kind of duality between different physical theories arises. The language of the two time physics is quite adapted also for the description of  field theories \cite{Bars7,Bars8} and of the gravity \cite{Bars9,Bars10}. A new approach to cosmology, inspired by two time physics has open the way to an interesting treatment of the problem of passing through the cosmological singularities \cite{Bars94,Bars95,Bars96,Bars97,Bars98}. 

An attractive feature of the two time physics is the fact that it permits to unify the systems and phenomena which before looked quite unconnected. How does it work? The approach is based on the duality between generalized coordinates and their conjugate momenta in any theory defined in the phase space. Indeed, if one  makes a linear transformation between the coordinates and momenta, one can require that this transformation  leave the Poisson brackets (or the commutators 
in quantum theory) intact. It is easy to check that the special linear transformations (i.e. transformations with the determinant equal to 1) satisfy this condition, which is both necessary and sufficient. Thus, there is the symmetry with respect to the group 
$SL(2,\mathbb R)$. We shall call here this group $Sp(2,\mathbb R$).  It can be called also $SO(1,2)$ and in the Bianchi classification it has the name of the Bianchi - VIII group \cite{Bianchi}. We can try to localize this group and its parameters become time-dependent or spacetime dependent. Then the symmetry disappears and we have to introduce gauge fields. 
The appearance of the gauge fields  is equivalent to the introduction of the first-class constraints and, as a consequence, one has to make a  choice of the gauge-fixing relations (see e.g. \cite{constr}). It was shown (see \cite{Bars-book} and  references therein) 
that to supply the theory with some physical degrees of freedom and at the same time avoid the arising of ghosts, it is necessary to add to the standard spacetime one additional time dimension and one additional space dimension. Different choices of the gauge-fixing conditions in this $d+2$ dimensional spacetime correspond to different choices of the time parameter, of the Hamiltonian and, generally, of the physical system in the standard $(d-1)+1$ dimensional spacetimes. In this way 2T physics arises.

As was already mentioned above in this framework some particular choices of the gauge-fixing produce the Hamiltonians of the non-relativistic particle, relativistic massive and massless particles and so on. In the present paper we shall make an attempt 
to explore the opportunity to describe  Carroll particles in the framework of the two time physics. Carroll particles are particles living in the spacetime, where the geometry is invariant with respect to the Carroll group. This group was introduced in 1965 by J.-M. L\'evy-Leblond \cite{LL}, who has noticed that applying the Wigner-In\"onu contraction \cite{WI} to the Poincare group one can obtain non only the Galilei group, corresponding to the infinite velocity of light, but also another group, arising in the limit when the velocity of light is equal to zero. He has called this group ``Carroll group'' as a tribute to Lewis Carroll books about Alice.  Independently the Carroll group was discovered by  N. D. Sen Gupta \cite{Gupta} in 1966. The properties of the large set of kinematic Lie algebras, including the Carroll Lie algebra were studied in \cite{LL-B}. For decades the Carroll group was considered as some bizzare mathematical curiosity, but now the interest to it is growing and a lot of quite unexpected applications are discovered \cite{Duval0,Duval,Duval1,Bergshoeff,Ciambelli,Donnay,Henneaux,Boer,Gomis,Hansen,Campoleoni,Figueroa,Figueroa1,Boer1}(for the history of the Carroll group see a recent paper by J.-M. L\'evy-Leblond \cite{LL1}).   

Let us note that first of all the Carroll group and the corresponding geometry are very interesting from the mathematical point of view \cite{Ciambelli,Gomis,Figueroa}. Indeed, one can study the intricate relations between the Poincare, Galilei, Carroll and Bargmannian \cite{Bargmann} Lie algebras (see \cite{Figueroa} and references therein). 
There is also an interesting relation with the infinite-dimensional Kac-Moody algebras \cite{Kac} (see \cite{Gomis}) and there is also a connection between the Carroll group and the Bondi-Metzner-Sachs group \cite{BMS,BMS1} (see \cite{Ciambelli}).  Amongst physical applications of the Carroll group we can mention an opportunity to construct the perturbative theory analogous to the post-Newton approximation in General Relativity, namely, one can consider the situations when the velocities of the objects under consideration are comparable with the velocity of light \cite{Hansen}.  Another interesting situation arises in cosmology when the recession velocity of distant objects in the universe is larger than the velocity of light. Such a situation also can be treated in Carrollian terms \cite{Boer}. Moreover, in the vicinity of the cosmological singularity where the oscillatory approach to the singularity takes place \cite{BKL} and where the evolutions in different spatial points become independent \cite{BKL1}, we again can speak about the correspondence between the Belinsky-Khalatnikov-Lifhshitz limit and the Carroll limit. The list  of applications is certainly much larger (see e.g. \cite{Boer1}).

The structure of our paper is the following: the rest of the Introduction contains two subsections which introduce briefly some useful notions and formulas for 2T physics and for the Carroll symmetry; the second section contains our version of classical theory of Carroll particles in 2T physics, while the third section is devoted to their quantization; the last section contains some concluding remarks. 

\subsection{Two time physics}

We have already mentioned that from the point of view of the 2T Physics,  usual physical systems living in a one time world represent  projections from the spacetime with one additional temporal dimension and one  additional spatial dimension. These additional dimensions are  introduced to construct a new gauge theory, based on the localization    of the phase-space symmetry described by the symplectic group $Sp(2,\mathbb{R})$. Then, the usual physics with 1T is obtained by means of a gauge fixing.
	
	In order to see how it works, let us introduce the phase-space coordinates for the two time world 
	\begin{gather}
		X^M=\left(X^{0'},X^{1'},X^\mu\right)\qquad P^M=\left(P^{0'},P^{1'},P^\mu\right).
	\end{gather}
	The indices $0'$ and $1'$ label an extra time and an extra space dimensions. We will see that the extra space dimension is necessary to get the right number of degrees of freedom in the 1T theory. The index $\mu=0,\dots ,d-1$ labels usual coordinates in one time world. We can \textit{pack} the position $X^M$ and the momentum coordinates $P^M$ as follows
	\begin{align}
		X^M_i=\left(X^M,P^M\right),
	\end{align}
	where $i=1,2$ labels mean the position and momentum respectively. In this way the two types of phase variables become indistinguishable and can be mixed  through  $Sp(2,\mathbb{R})$ transformations. 
	
	Now, let us consider the worldline action for a free particle in a flat two time spacetime
	\begin{gather}\label{WorldAction}
		S=\frac{1}{2}\int d\tau\ \epsilon^{ij}\eta_{MN}\ \partial_\tau X_i^M X_j^N,
	\end{gather}
	where ${\eta_{MN}=\text{Diag(-1,1,-1,1,\dots,1)}}$ is the flat metric, with signature $(2,d)$, and $\epsilon^{ij}$ is the antisymmetric tensor with $\epsilon^{12}=1$. $\tau$ is a proper time parameter that parametrizes the worldline action in the two time plane. Notice that the action \eqref{WorldAction} is invariant under the global $Sp(2,\mathbb{R})$ transformations, which infinitesimal form is given by
	\begin{align}
		\delta_\omega X_i^M = \epsilon_{ij}\omega^{jk}X_k^M.
	\end{align}
	Here,  the transformation parameters $\omega^{jk}$ are symmetric in $j,k$.
	
	What does  happen when the $Sp(2,\mathbb{R})$ symmetry is promoted to a local symmetry (in particular, when $\omega^{ij}\rightarrow\omega^{ij}(\tau)$)? In this case we need to introduce a connection that takes into account the new gauge symmetry. The derivative with respect to $\tau$ is substituted by  a covariant derivative
	\begin{align}
		\partial_\tau X^M_i\rightarrow D_\tau X^M_i=\partial_\tau X^M_i-\epsilon_{ij} A^{jk}(\tau)X^M_k,
	\end{align}
	where $A^{jk}(\tau)$ is symmetric in the indices $i,j$ and belongs to the adjoint representation of the  Lie algebra of $Sp(2,\mathbb{R})$ (that we call $\mathfrak{sp}(2,\mathbb{R})$). It transforms as a gauge field under the $Sp(2,\mathbb{R})$ group
	\begin{align}
		\delta_\omega A^{ij}(\tau)=\partial_\tau\omega^{ij}+\omega^{ik}\epsilon_{kl}A^{lj}+\omega^{jk}\epsilon_{kl}A^{li}.
	\end{align}
	The worldline action invariant under these gauge transformations is
	\begin{align}\label{WorldGaugedAction}
		S&=\frac{1}{2}\int d\tau\ \epsilon^{ij}\eta_{MN}D_\tau X_i^M X_j^N\cr
		&\qquad=\int d\tau\ \left[\eta^{MN} \partial_\tau X_MP_N-A^{ij}(\tau)Q_{ij}\right],
	\end{align}
	where
	\begin{gather}
		Q_{11}=\frac{1}{2}X\cdot X,\quad Q_{22}=\frac{1}{2}P\cdot P,\cr 
		Q_{12}=Q_{21}=\frac{1}{2}X\cdot P
	\end{gather}
	are the $\mathfrak{sp}(2,\mathbb{R})$ conserved currents. 
	%As described in \cite{Bars2008:Gravity,Bars2010:BBandInflation}, the $Q_{ij}$ takes more complicated forms when a non-flat metric is considered, but we will not discuss it in the present paper: for now, we will concentrate on the flat space systems.
	
	The gauge fields $A^{ij}$ are not dynamical, the kinetic terms for them are absent.  Thus, in the action \eqref{WorldGaugedAction} they play the role of Lagrange multipliers. When a gauge is chosen, the following constraints must be satisfied
	\begin{align}\label{CI}
		X\cdot X&=0,\\\label{CII}
		X\cdot P&=0,\\\label{CIII}
		P\cdot P&=0.
	\end{align}
	It is worth noticing that these constraints lead to a non-trivial parameterization of the 1T spacetime only when the starting theory has more than one timelike dimension (see, for instance, \cite{Bars3,Bars-book} and references therein).
	Moreover, the gauge freedom allows us to choose three physical degrees of freedom. Then, when the gauge is fixed and the constraints \eqref{CI}, \eqref{CII} and \eqref{CIII} are satisfied, one gets the right number of 1T variables. That means that $X^M_i(\tau)=X^M_i(\vec{x}(\tau),\vec{p}(\tau))$. The action now looks like
	\begin{align}
		S=\int d\tau \left(\dot{\vec{x}}\cdot\vec{p}-H\right),
	\end{align}
	where $H$ is the Hamiltonian of the 1T theory.
	
	What is the meaning of this gauge fixing? Different gauge fixings correspond to different choices of the Hamiltonian (and different choices of the time). This means that different systems in the 1T physics are described by a unique two time model. In this sense, these systems are \textit{dual} to each other under local $Sp(2,\mathbb{R})$ transformations. 
	
	In many cases it is useful to fix the gauge \textit{partially}. For instance, one can make two gauge choices and solve two of the constraints \eqref{CI}, \eqref{CII} and \eqref{CIII}. This way, the remaining quantities will be written in terms of the phase-space variables $(t,\vec{x},H,\vec{p})$. The last gauge fixing will set the physical time $t$ in terms of $\tau$ (this corresponds to a \textit{choice of the time}) and will define the Hamiltonian $H$. For example, we may fix $X\cdot X=X\cdot P=0$. At this stage, the action takes the form
	\begin{align}
		S=\int d\tau \left(\dot{\vec{x}}\cdot\vec{p}-\dot{t}H-\frac{A^{22}}{2}P\cdot P\right).
	\end{align}
	The last constraint, defined by $P\cdot P=0$, generally characterizes the theory we want to describe. For instance, if we are describing a massless relativistic particle, $P\cdot P = p^2$, where $p$ is the quadrimomentum, or for a non-relativistic particle, $P\cdot P=\vec{p}^2-2mH$.
	
	Let us take a closer look at the equations of motion and the symmetries of the system. 
	
	The equations of motions for the action \eqref{WorldGaugedAction} are
	\begin{align}\label{EqI}
		\dot{X}^M&=A^{12}X^M+A^{22}P^M\\\label{EqII}
		\dot{P}^M&=-A^{12}P^M-A^{22}X^M.
	\end{align}
	Thus, now we see how the choice of the gauge fields $A^{ij}$ affects the equations of motion. The equations and the action \eqref{WorldGaugedAction} are invariant under the gauge group $Sp(2,\mathbb{R})$ and the global transformations $SO(2,d)$, where $d$ is the  dimension of  our one-time spacetime. This group plays an  important role in both classical and quantum theory. The generators of the group $SO(2,d)$ are 
	\begin{align}
		L^{MN}=X^MP^N-X^NP^M,
		\label{generator}
	\end{align}
	and are invariant under $Sp(2,{\mathbb R})$ transformations. When the gauge is fixed partially, these generators are written in terms of the 1T variables and a subset of them will provide the generators of the symmetries of the 1T subsystem. 
	%It has to be mentioned here the the remaining generators do are not symmetries of the equations of motion, but they always describe the symmetries of the action. More in general, they represent the conformal transformations of the system. A relevant result in this direction has been obtained, for instance, for the Hydrogen atom \cite{Bars1998:FreeArmonicHAtom}, in which the group $SO(2,4)$ includes the so-called \textit{hidden symmetries} of the system, which arise quite naturally from the 2T Physics (see also \cite{Bars1999:NonRelMass}).
	
	%In order to fix these ideas, the reader may refer to the work in \cite{Bars2007:DualTheories}, in which some examples and further readings are provided.
	Let us finish this subsection with the note that it was based mainly on papers \cite{Bars3,Bars4,Bars6}.
\\
\subsection{Carroll symmetry}

It is well known that the Poincar\'e group possesses the contraction, obtained by sending the speed of light to infinity $c\rightarrow\infty$. This limit leads to the Galilean group that describes non-relativistic models. Indeed, the transformations of the space-time coordinates, given by the Poincar\'e group are
	\begin{align}\label{LotentzTransf}
		\left\{
		\begin{array}{l}
			x'_0=\gamma\left[x_0+\vec{\beta}\cdot(R\vec{x})\right]+a_0,\\
			\vec{x}'=R\vec{x}+\frac{\gamma^2}{\gamma+1}\left[\vec{\beta}\cdot(R\vec{x})\right]\vec{\beta}+\gamma\vec{\beta}x_0+\vec{a}.
		\end{array}
		\right.
	\end{align}
	Here, $(x_0,\vec{x})$ are the time and the space coordinates respectively. Then, we can recognize the rotation transformation $R$, the time and space translations $(a_0,\vec{a})$ and the boost parameters $\vec{\beta}=\vec{v}/c$ and $\gamma=1/\sqrt{1-(v/c)^2}$, where $\vec{v}$ is the boost velocity and $c$ is the speed of light. The non-relativistic limit is obtained by the transition to  the new variables
	\begin{align}
		t=\frac{1}{c}x_0\quad \vec{v}=c\vec{\beta}\quad b=\frac{1}{c}a_0.
	\end{align}
	When the speed of light is sent to infinity, these quantities remain constant. This means that the transformations \eqref{LotentzTransf} reduce to
	\begin{align}
		\left\{
		\begin{array}{l}
			t'=t+b,\\
			\vec{x}'=R\vec{x}+\vec{v}t+\vec{a}.
		\end{array}
		\right.
	\end{align}
	That is exactly what we expect from the Galilean transformations.
	
	On the other hand, we can also wonder what does  happen when we consider the opposite limit; namely, for $c\rightarrow 0$. To see it, let us define the new variables
	\begin{align}
		t=\frac{1}{c}x_0\quad \vec{\hat{v}}=\frac{1}{c}\vec{\beta},\quad b=\frac{1}{c}a_0,
	\end{align}
	requiring that they remain constant after $c$ is sent to zero. We get the following transformations
	\begin{align}
		\left\{
		\begin{array}{l}
			t'=t+\vec{v}\cdot(R\vec{x})+b,\\
			\vec{x}'=R\vec{x}+\vec{a}.
		\end{array}
		\right.
	\end{align}
	These transformations form the Carroll group \cite{LL,Gupta,LL-B}. Note that here the boosts transform only the time coordinate, in contrast to what happens at the Galilean transformations. In the Table \ref{Tab:Algebras} we present the nonvanishing commutators of the generators of all three groups: Poincar\'e, Gallilei and Carrol. 
	%This limit has some consequences also in the algebra of the generators of the symmetries. As its known, the symmetries in the relativistic theory is given by the Poincare group. The limits for $c\rightarrow\infty$ and $c\rightarrow 0$ lead to the Galileo group and Carroll group respectively. The algebra of the generators is summarized in Table \ref{Tab:Algebras}, where the non-zero lie brackets for the Poincare algebra and their corresponding values after the two limits are reported.
	\begin{table*}[!tbh]
		\begin{tabular}{ll|ll}
			\hline
			\multicolumn{4}{c}{POINCAR\'E ALGEBRA}\\
			\hline
			\multicolumn{4}{c}{\vspace{-10pt}}\\
			\qquad\qquad\qquad\qquad\qquad\qquad&\multicolumn{2}{l}{$[L^{ij},L^{kl}]=\delta^{ik}L^{jl}+\delta^{jl}L^{ik}-\delta^{il}L^{jk}-\delta^{jk}L^{il}$}& \\
			\qquad\qquad\qquad\qquad\qquad\qquad&\multicolumn{2}{l}{$[L^{ij},P^k]=\delta^{ik}P^j-\delta^{jk}P^{i}$}& \\
			\qquad\qquad\qquad\qquad\qquad\qquad&\multicolumn{2}{l}{$[L^{ij},B^k]=\delta^{ik}B^j-\delta^{jk}B^{i}$}& \\
			\qquad\qquad\qquad\qquad\qquad\qquad&\multicolumn{2}{l}{$[B^i,B^j]=L^{ij}$}& \\
			\qquad\qquad\qquad\qquad\qquad\qquad&\multicolumn{2}{l}{$[B^i,P^j]=\delta^{ij}H$}& \\
			\qquad\qquad\qquad\qquad\qquad\qquad&\multicolumn{2}{l}{$[B^i,H]=P^i$}& \\
			\multicolumn{4}{c}{\vspace{1pt}}\\
			\hline
			\multicolumn{2}{c|}{GALILEI ALGEBRA}&\multicolumn{2}{|c}{CARROLL ALGEBRA}\\
			\hline
			\multicolumn{2}{c|}{\vspace{-2pt}}&\multicolumn{2}{|c}{\vspace{-7pt}}\\
			\multicolumn{2}{l|}{$[L^{ij},L^{kl}]=\delta^{ik}L^{jl}+\delta^{jl}L^{ik}-\delta^{il}L^{jk}-\delta^{jk}L^{il}$} &  \multicolumn{2}{|l}{$[L^{ij},L^{kl}]=\delta^{ik}L^{jl}+\delta^{jl}L^{ik}-\delta^{il}L^{jk}-\delta^{jk}L^{il}$} \\
			\multicolumn{2}{l|}{$[L^{ij},P^k]=\delta^{ik}P^j-\delta^{jk}P^{i}$}&\multicolumn{2}{|l}{$[L^{ij},P^k]=\delta^{ik}P^j-\delta^{jk}P^{i}$} \\
			\multicolumn{2}{l|}{$[L^{ij},B^k]=\delta^{ik}B^j-\delta^{jk}B^{i}$}&\multicolumn{2}{|l}{$[L^{ij},B^k]=\delta^{ik}B^j-\delta^{jk}B^{i}$} \\
			\multicolumn{2}{l|}{$[B^i,B^j]=0$}&\multicolumn{2}{|l}{$[B^i,B^j]=0$} \\
			\multicolumn{2}{l|}{$[B^i,P^j]=0$}&\multicolumn{2}{|l}{$[B^i,P^j]=\delta^{ij}H$} \\
			\multicolumn{2}{l|}{$[B^i,H]=P^i$}&\multicolumn{2}{|l}{$[B^i,H]=0$}
		\end{tabular}
		\caption{Lie brackets for the Galilei, Poincar\'e and Carroll algebras. Here, $L^{ij}$ represent the generators of rotations, $B^i$ are the boosts and $H$ and $P^i$ generates the time and space translations respectively.}
		\label{Tab:Algebras}
	\end{table*}
	
	Let us see what happens with the metric tensor. The Lorentz metric is 
	\begin{align}
		ds^2=-c^2\ dt^2 + d\vec{x}^2.
	\end{align}
	When the Galilean or Carrollian limits are considered, the contravariant metric (in the Galilean case) or the metric (in the Carrollian case) becomes degenerate 
	(see e.g. \cite{Figueroa}).
	%\cite{JMLL-CarrollOriginal,JMLL:PossibleKinematics,Bergshoeff:CarrollVSGalilei,Ciambelli:CarrollStructures}. 
It is interesting to represent it in terms of the light cone, which determines the causal structure of the spacetime. 
	
	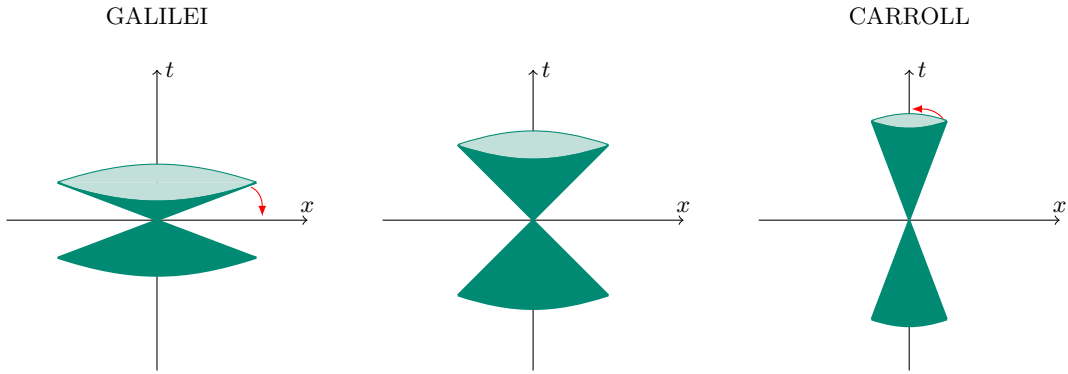
\begin{figure*}[!hpbt]
		\begin{tikzpicture}
			\draw [->] (-7,0) -- (-3,0) 
			node [anchor=south] {$x$};
			\draw [->] (-5,-2) -- (-5,2) 
			node [anchor=west] {$t$}
			node[above,yshift=.5cm]{GALILEI};
			\draw[PineGreen,thick](-3.68,0.5) to[out=162,in=18] (-6.32,0.5);
			\fill [PineGreen!20,nearly transparent](-3.68,0.5) to[out=162,in=18] (-6.32,0.5);
			\draw[PineGreen,thick](-3.68,0.5) to[out=-162,in=-18] (-6.32,0.5);
			\fill [PineGreen!20,nearly transparent](-3.68,0.5) to[out=-162,in=-18] (-6.32,0.5);
			\draw[PineGreen,dashed,thick](-3.68,-0.5) to[out=162,in=18] (-6.32,-0.5);
			\draw[PineGreen,thick](-3.68,-0.5) to[out=-162,in=-18] (-6.32,-0.5);
			\draw [draw=PineGreen, thick, text=blue] (-6.32,-0.5)--(-3.68,0.5);
			\draw [draw=PineGreen, thick, text=blue] (-6.32,0.5)--(-3.68,-0.5);
			\draw [->] (-2,0) -- (2,0) 
			node [anchor=south] {$x$};
			\draw [->] (0,-2) -- (0,2) 
			node [anchor=west] {$t$};
			\draw[PineGreen,thick](1,1) to[out=162,in=18] (-1,1);
			\fill [PineGreen!20,nearly transparent](1,1) to[out=162,in=18] (-1,1);
			\draw[PineGreen,thick](-1,1) to[out=-18,in=-162] (1,1);
			\fill [PineGreen!20,nearly transparent](-1,1) to[out=-18,in=-162] (1,1);
			\draw[PineGreen,dashed,thick](1,-1) to[out=162,in=18] (-1,-1);
			\draw[PineGreen,thick](1,-1) to[out=-162,in=-18] (-1,-1);
			\draw [draw=PineGreen, thick, text=blue] (-1,-1)--(1,1);
			\draw [draw=PineGreen, thick, text=blue] (-1,1)--(1,-1);
			\draw [->] (3,0) -- (7,0) 
			node [anchor=south] {$x$};
			\draw [->] (5,-2) -- (5,2) 
			node [anchor=west] {$t$}
			node[above,yshift=.5cm]{CARROLL};
			\draw[PineGreen,thick]((5.5,1.32) to[out=162,in=18] (4.5,1.32);
			\fill [PineGreen!20,nearly transparent]((5.5,1.32) to[out=162,in=18] (4.5,1.32);
			\draw[PineGreen,thick](4.5,1.32) to[out=-18,in=-162] ((5.5,1.32);
			\fill [PineGreen!20,nearly transparent](4.5,1.32) to[out=-18,in=-162] ((5.5,1.32);
			\draw[PineGreen,dashed,thick](5.5,-1.32) to[out=162,in=18] (4.5,-1.32);
			\draw[PineGreen,thick](5.5,-1.32) to[out=-162,in=-18] (4.5,-1.32);
			\draw [draw=PineGreen, thick, text=blue] (4.5,-1.32)--(5.5,1.32);
			\draw [draw=PineGreen, thick, text=blue] (4.5,1.32)--(5.5,-1.32);
			\draw[-latex,red] (5.45,1.35) [out=130,in=0] to (5.03,1.48);
			\draw[-latex,red] (-3.75,0.44) [out=-25,in=90] to (-3.6,0.05);
			\fill[PineGreen,nearly transparent](-5,0)--(-3.68,0.5) to[out=-162,in=-18] (-6.32,0.5)--cycle;
			\fill[PineGreen,nearly transparent](-5,0)--(-3.68,-0.5) to[out=-162,in=-18] (-6.32,-0.5)--cycle;
			\fill[PineGreen,nearly transparent](0,0)--(1,1) to[out=-162,in=-18] (-1,1)--cycle;
			\fill[PineGreen,nearly transparent](0,0)--(-1,-1) to[out=-18,in=-162] (1,-1)--cycle;
			\fill[PineGreen,nearly transparent](5,0)--(5.5,1.32) to[out=-162,in=-18] (4.5,1.32)--cycle;
			\fill[PineGreen,nearly transparent](5,0)--(4.5,-1.32) to[out=-18,in=-162] (5.5,-1.32)--cycle;
		\end{tikzpicture}
		\caption{Deformation of the Lorentzian light-cone in Galilean and Carrolian limits.}
		\label{Fig:LC}
	\end{figure*}
	
	In Fig.\ref{Fig:LC}  the light cones for the Lorentzian, Galilean and Carrollian spaces are presented. The light cone becomes a degenerate timelike line in the Carrollian case and a spacelike hyperplane in the Galilean case. It is known that the latter describes non-relativistic theories. On the other hand, the first picture tell us that the Carroll particle cannot move in space unless it is a \textit{tachyon}. This situation is  described, for example, in \cite{Boer}. In this paper  our goal is to describe the non-moving  Carroll particles in the 2T world.  
\\	
\section{Classical theory}

It is known that the Carrol particle with non-zero energy should be always in rest \cite{Duval,Bergshoeff,Boer}. In the paper \cite{Boer} it was explained that the Carroll particle
with zero  energy  is instead always moving. These two cases are not connected. Indeed, the Carroll boosts do not change the value of the energy in contrast to the Lorentzian and Galilean boosts. It is easy to understand the reason. The Lie algebra generators corresponding to the Lorentz boosts have the form
\begin{equation}
t\frac{\partial}{\partial x} + x\frac{\partial}{\partial t},
\label{Lor}
\end{equation}
the Galilean boosts can be represented by the generators
\begin{equation}
t\frac{\partial}{\partial x},
\label{Gal}
\end{equation}
while the Carroll boosts are
\begin{equation}
x\frac{\partial}{\partial t}.
\label{Car}
\end{equation} 
The Hamiltonian is always proportional to the operator 
\begin{equation}
\frac{\partial}{\partial t}.
\label{Ham}
\end{equation}
The Carroll boost \eqref{Car} commutes with the Hamiltonian \eqref{Ham} in contrast to the Lorentz boost \eqref{Lor} and Galilei boost \eqref{Gal}.
Thus, one cannot change the value of the energy making a boost in the Carroll world. Hence, one should treat the cases of the vanishing and non-vanishing energy separately. In this paper we discuss only the case of the Carroll particles with non-vanishing energy.   	
\\
%\subsection{Particle at rest}

The fact that in the Carroll spacetime particles with nonzero energy cannot move can be  explained also in the following way \cite{Boer}:
the conservation of the energy-momentum tensor implies the disappearance of the flux of energy, if the energy is different from zero. Then, following the paper \cite{Boer} one states that 
the Carrollian particle's dynamics can be described by the following action in the first order formalism
	\begin{align}\label{CarrollActionRest}
		S=-\int d\tau\left\{\dot{t}E-\dot{x}\cdot p - \lambda\left(E-E_0\right) \right\},
	\end{align}
	where $\tau$ is the proper time, $t$ is the physical time, $E$ represents the classical Hamiltonian and $x^i$ and $p^i$ are the space coordinates and the momenta, for $i=1,\dots, d-1$. The usual scalar product is defined by $x\cdot p=\sum_{i,j} \eta_{ij}x^ip^j$; here $\eta_{ij} = \delta_{ij}$. 
	Besides,  $E_0$ is a real constant that represents the rest energy of the Carroll particle and $\lambda$ plays the role of a Lagrange multiplier. The dot represents the derivative with respect to $\tau$. This action is invariant under the transformations generated by
	\begin{align}
		L^{ij}=x^{i}p^{j}-x^jp^i,\quad B^i=Ex^i, \quad p^i\  {\rm and}\  E,
	\end{align}
	which are respectively the generators of rotations, boosts and spatial and time translations. Their Poisson brackets satisfy the Carroll algebra \cite{LL,Gupta,LL-B}. The equations of motion following from the action \eqref{CarrollActionRest} are 
	\begin{gather}
		\dot{t}=\lambda,\qquad\dot{E}=0,\cr \dot{x}^i=0,\ \ \dot{p}^i=0.
	\end{gather}
	In particular, the first equation fixes   the relation between the Lagrange multiplier, the proper time and the physical time: from $\lambda=1$, it follows that $t=\tau$.
	
	Let us come back  to the two time physics. We can suppose  that the gauge fields $A^{ij}$ are all constants. In particular, we choose $A^{12}=A^{11}=0$. Here, it is worth observing that if $X^i=x^i$ and $P^i=p^i$, it requires
	\begin{align}
		\dot{X}^i=0,\ \  \dot{P}^i=0.
	\end{align}
	Combined with  equations \eqref{CI}, \eqref{CII} and \eqref{CIII}, this leads to the condition $x^i\propto p^i$. This is a very strong bound, which looks too restrictive and rather unnatural. To avoid the imposing of this condition we can choose a more involved relation between the physical coordinates and momenta in our one time spacetime and the coordinates and momenta in the two time spacetime:
	
	\begin{align}\label{CoordChoice}
		X^M=X^M_0+tP^M_0,\ \ \  P^M=P^M_0,
	\end{align}
	where $X^M_0$ and $P^M_0$ are independent on $t$. It is important to observe that here $t$ has not been defined in terms of $X^M$ and $P^M$ yet. We are going to define it later, but before let us see the consequences of our choice \eqref{CoordChoice}.
	
	The action obtained by imposing  $A^{11}=A^{12}=0$ in \eqref{WorldGaugedAction} is
	\begin{align}
		S=\int d\tau\left[\dot{X}\cdot P-\frac{1}{2}A^{22}\ P\cdot P\right].
	\end{align}
	It is clear that if we want to reproduce the action \eqref{CarrollActionRest}, we have to require that $P\cdot P=-2(E-E_0)$. Now  the gauge field $A^{22}$ plays the role of  the Lagrange multiplier $\lambda$.
	
	When $M=i$, we can simply define $X^i_0=x^i$ and $P^i_0=p^i$, where $x^i$ and $p^i$ are the position and momentum in the 1T theory. Supposing that the equations $\dot{x}^i=\dot{p}^i=0$ are satisfied, the equations \eqref{EqI} and \eqref{EqII} for $X^i=x^i+tp^i$ and $P^i=p^i$ become
	\begin{align}
		\dot{X}^i=\dot{t}p^i=A^{22}p^i,\ \ \  \dot{P}^i=0,
	\end{align}
	which are satisfied when $\dot{t}=A^{22}$, as expected.
	
	Now using our coordinate choice \eqref{CoordChoice} and the constraints \eqref{CI} and \eqref{CII} we find that 
	\begin{align}
		X\cdot X&=-2X_0^+X_0^--(X_0^0)^2+x^ix_i\cr &\quad+2t\left(-X^+_0P^-_0-X^-_0P^+_0-X^0_0P^0_0+p^ix_i\right)\cr &\qquad+t^2\left[-2P^+_0P^-_0-(P^0_0)^2+p^ip_i\right],\\
		X\cdot P&=-X_0^+P_0^--X_0^-P_0^+-X_0^0P_0^0+x^ip_i\cr &\quad+t\left[-2P_0^+P^-_0-(P^0_0)^2+p_ip^i\right].
	\end{align}
	Here, we can set
	\begin{align}\label{ConstI}
		&-2X_0^+X_0^--(X_0^0)^2+x^ix_i=0,\\\label{ConstII}
		&-X^+_0P^-_0-X^-_0P^+_0-X^0_0P^0_0+p^ix_i=0.
	\end{align}
	Now, it is time to remember that we need
	\begin{align}\label{ConstIII}
		P\cdot P=-2P_0^+P^-_0-(P^0_0)^2+p_ip^i=-2(E-E_0).
	\end{align}
	This last condition is treated not  as a constraint, but as  the definition of the function
	\begin{align}
		E=\frac{1}{2}\left(2P_0^+P^-_0+(P^0_0)^2-p_ip^i\right)+E_0,
	\end{align}
	where $E_0$ is a constant. When these equations are solved, we are left with
	\begin{align}
		P\cdot P&=-2(E-E_0)\\ 
		X\cdot P&=-2t(E-E_0)\\ 
		X\cdot X&=-2t^2(E-E_0).
	\end{align}
	This means that all these constraints are satisfied if the  same equality is valid. Namely,
	\begin{align}
		X\cdot X=tX\cdot P=t^2P\cdot P,
	\end{align}
	which are satisfied if and only if  $E-E_0=0$.
	
	After this preliminary analysis we can represent the procedure of the reduction of the two time spacetime to the one time spacetime in a consistent way following the algorithm presented in papers \cite{Bars3,Bars4,Bars6}.  
	Using our coordinate (gauge-fixing) choice we can calculate the kinetic term of the action \eqref{CarrollActionRest} which now looks like 
	\begin{align}
		\dot{X}\cdot P&=\dot{X}_{0}\cdot P_0+\dot{t}P_{0}\cdot P_0+t\dot{P}_{0}\cdot P\cr 
		&=\dot{X}_{0}\cdot P_0+\frac{1}{2}\dot{t}P_{0}\cdot P_0+\partial_\tau(tP_{0}\cdot P).
	\end{align}
	Using the definition \eqref{ConstIII}, we can write
	\begin{align}
		\dot{X}\cdot P=\dot{X}_{0}\cdot P_0-\dot{t}E+(\text{a total derivative}).
	\end{align}
	Now the action can be  written as
	\begin{align}\label{ActionSemifinalStep}
		S&=\int d\tau\Bigg\{ 
		-\dot{X}^+_0P^-_0-\dot{X}^-_0P^+_0-\dot{X}^0_0P^0_0
		\cr 
		&\qquad\qquad\quad+\dot{x}^ip^i-\dot{t}E+A^{22}\left(E-E_0\right)
		\Bigg\}.
	\end{align}
	In order to obtain the action \eqref{CarrollActionRest}, it is necessary to combine the first three terms in the integral above into  a total derivative. We can use our gauge freedom in order to reach this goal.  Moreover, it is important to remember that the function $t$ has not been defined in terms of $X^M$ and $P^M$ yet, but it is always an additional degree of freedom: we \textit{need} to set the $X_0^M$ and $P_0^M$ in order to define $t$. Therefore, we are free to fix three coordinates in the two time space, paying attention not to spoil the contraints \eqref{ConstI}, \eqref{ConstII} and \eqref{ConstIII}. Our choice is summarized in Table \ref{Tab:RestParticle}, where the physical time is  defined as $t=\frac{X^+}{E_0}$. In this way, the action \eqref{ActionSemifinalStep} becomes
	\begin{align}\label{CarrollActionTrovata}
		S=-\int d\tau\Bigg\{
		\dot{t}E-\dot{x}_ip^i-A^{22}\left(E-E_0\right),
		\Bigg\},
	\end{align}
	which coincides with the Carroll action \eqref{CarrollActionRest}.
	
	\vspace{5pt}
	\begin{table*}[!hpbt]
		\centering\renewcommand{\arraystretch}{1.5}
		\scalebox{1.2}{
		\begin{tabular}{ c | c  c  c  c }
			& $+\qquad$ & $-\qquad$ & $0\qquad$ & $i$ \\
			\hline
			$X^M$ & $E_0\ t\qquad$ & $\frac{p^ix_i}{E_0}+\frac{t}{E_0}\left(E-E_0+\frac{p_ip^i}{2}\right)\qquad$ & $\sqrt{x_ix^i}\qquad$ & $x^i+tp^i$ \\
			$P^M$ & $E_0\qquad$ & $\frac{E-E_0+\frac{p_ip^i}{2}}{E_0}\qquad$ & $0\qquad$ & $p^i$\\
		\end{tabular}}
		\caption{Gauge-fixing choice for the Carroll  particle with non-vanishing energy.}
		\label{Tab:RestParticle}
	\end{table*}
	\vspace{5pt}
	
	%\subsubsection{The algebra of the generators of $SO(2,d)$ for the particle at rest}
	We can now compute the $SO(2,d)$ generators in terms of the variables defined by  our gauge choice. Its general form was given in Eq. \eqref{generator}. 
%	\begin{align}
%		L^{MN}=X^MP^N-X^NP^M.
%	\end{align}
	Using the gauge choices defined above, we obtain the following expressions: 
	\begin{align}\label{S:GenRowI}
		L^{ij}&=x^{i}p^{j}-x^{j}p^i,\\ \label{S:GenRowII}
		L^{0i}&=\sqrt{x^jx_j}\ p^i,\\ \label{S:GenRowIII}
		L^{+i}&=-E_0 x^i\\\label{S:GenRowIV} L^{-i}&=-\frac{E-E_0}{E_0}x^i-\frac{p_jp^j}{2E_0}x^i+\frac{p^jx_j}{E_0} p^i,\\\label{S:GenRowV}
		L^{+-}&=-p^ix_i,\\\label{S:GenRowVI}
		L^{-0}&=-\sqrt{x_ix^i}\left(\frac{E-E_0}{E_0}+\frac{p_ip^i}{2E_0}\right),\\ \label{S:GenRowVII}
		L^{+0}&=-E_0\sqrt{x_ix^i}.
	\end{align}
	Using the definition of the canonical Poisson brackets 
	\begin{align}
		\{x^i,p^j\}=\eta^{ij},
	\end{align}
	one can determine the infinitesimal transformation of the phase space coordinates under the group generated by \eqref{S:GenRowI}-\eqref{S:GenRowVII}:
	\begin{widetext}
		\begin{align}
			\delta x^k&=\epsilon^{ki}x_i+\epsilon^{0k}r+\frac{\epsilon_{-i}}{E_0}\left(x^ip^k-x^kp^i-g^{ik}p\cdot x\right)+\epsilon_{+-}x^k+\epsilon_{-0}\frac{rp^k}{E_0}\\
			\delta p^k &= \epsilon^{ki}p_i+\epsilon_{0i}\frac{p^ix^k}{r}+\frac{\epsilon_{-i}}{E_0}\left[p^ip^k-g^{ik}\frac{p^2}{2}-g^{ik}(E-E_0)\right]-\epsilon_{+-}p^k\cr &\qquad\qquad\qquad\qquad\qquad-\frac{\epsilon_{-0}}{E_0}\left[\frac{p^2}{2}+(E-E_0)\right]\frac{x^k}{r}+E_0\left(\epsilon^{-k}-\epsilon_{+0}\frac{x^k}{r}\right),
		\end{align}
	\end{widetext}
	where the $\epsilon_{MN}$ are infinitesimal transformation parameters which correspond to the respective generators $L^{MN}$.   
	The action \eqref{CarrollActionTrovata} is invariant under these transformations, but the Poisson brackets of these generators do not form an $\mathfrak{so}(2,d)$ algebra, unless the constraint $E-E_0=0$ is satisfied. Indeed, a direct computation shows that
	\begin{align}
		\{L^{-i},L^{-j}\}=-2\frac{E-E_0}{E_0}L^{ij},
	\end{align}
	which is a new element of the algebra. On the other hand, when $E-E_0=0$ these Poisson brackets vanish and all the generators form the $\mathfrak{so}(2,d)$ algebra, described by
	\begin{align}
		&\{L^{MN},L^{RS}\}\cr 
		&\quad=\eta^{MR}L^{NS}+\eta^{NS}L^{MR}\cr &\qquad\qquad\qquad-\eta^{MS}L^{NR}-\eta^{NR}L^{MS}.
	\label{class-alg}
	\end{align}
	It is interesting to observe that the generators $L^{+i}$ can now be interpreted as the Carrollian boosts generators.
	Let us mention also the fact that we do not need to introduce the gauge transformation of the gauge field $A^{22}$ in contrast to the situation which was encountered, for example, for the non-relativistic particle \cite{Bars6}. 
	 \\

\section{Quantum Theory}

Now, we would like to present the results of the quantization of the model of the Carroll particle arising as the consequence  of the gauge choice in 2T spacetime described in the preceding section. Firstly, we recapitulate the general ideas following the papers \cite{Bars3,Bars4}. The first step consists in the definition of the commutation relation for the position and momentum operators in the standard $d-1$-dimensional space. They are, naturally.
	\begin{align}\label{RQCommRules}
		[x^i,p^j]=i\ \delta^{ij}.
	\end{align}
%	where $x^i$ and $p^j$ are the physical quantities. As also outlined in [CITE], the main difficulty resides in the resolution of the ordering ambiguities. Indeed, there are infinite ways in which an operator can be ordered, and each way change from the others by a commutator. Let us consider as an example the operator $p^2r$, which will be also useful in the following sections. We want to order it in such a way to be hermitian. It cab be written as
When we should quantize some classical functions of these operators, the problem of the choice of the ordering arises.
Firstly,  all the operators should be Hermitian, but this requirement is not sufficient. Let us consider as a simple example (arising also in \cite{Bars4}) the function $p^2r$, where $r = \sqrt{x_ix^i}$. We can write, for example,  
	\begin{align}\label{ReorderingI}
		p^2r\rightarrow p_irp^i,
	\end{align}
	which is clearly Hermitian. However, this ordering is not unique ordering providing the Hermiticity. Indeed,
	we can choose another form of the operator
	\begin{align}\label{ReorderingII}
		p^2r\rightarrow rp_ir^{-1}p^ir=p_irp^i-\frac{d-3}{2r},
	\end{align}
	where the right-hand side of the equation is  computed by means of the commutation rules \eqref{RQCommRules}. Surely both the expressions \eqref{ReorderingI} and \eqref{ReorderingII}  are  admissible. In a more general case, one obtains
	\begin{align}\label{ReorderingIII}
		p^2r\rightarrow r^\lambda p_ir^{1-2\lambda}p^ir^\lambda=p_irp^i+\frac{\lambda(\lambda-d+2)}{2r}.
	\end{align}

Thus, to make an adeguate choice of the ordering we have to resort to the covariant quantization in  the 2T spacetime. 
First of all the generators $L^{MN}$ which become operators should constitute the Lie algebra with respect to the commutators. This algebra should coincide with the Lie algebra \eqref{class-alg}  of the classical generators with respect to the Poisson bracket. However, this requirement also does not define the ordering in the quantum generators in a unique way and one should use also the properties of the Casimir operators of the unitary representations of both the groups  $SO(2,d)$ and $Sp(2,{\mathbb R})$. The constraints play the role of the generators of the symmetry with respect to the $Sp(2,{\mathbb R})$ group. 
Hence, in the quantum theory they should eliminate the acceptable quantum states of the system under consideration according to the prescription of the Dirac quantization of systems with first-class constraints (see, e.g. \cite{constr}). 
Then, if the generators eliminate the quantum states the same should be valid also for the Casimir operators. 
If we choose  the basis of the Hermitian quantum generators of  the $Sp(2,\mathbb R)$ group as follows (see \cite{Bars3}):
\begin{eqnarray}
&&J_0=\frac14(P^2+X^2),\ J_1 = \frac14(P^2-X^2),\nonumber \\
&& J_2 = \frac14(X\cdot P+ P\cdot X), 
\label{basis-Cas}
\end{eqnarray} 
we obtain the following commutation rules:
\begin{eqnarray}
&&[J_0,J_1]=iJ_2,\ [J_0,J_2] = -iJ_1,\nonumber \\
&&[J_1,J_2] = -iJ_0.
\label{basis-Cas1}
\end{eqnarray}
The quadratic Casimir operator is defined as 
\begin{equation}
C_2(Sp(2,\mathbb R) = J_0^2-J_1^2-J_2^2.
\label{basis-Cas2}
\end{equation}
Using the commutation rules 
\begin{equation}
[X^M,P^N]=i\eta^{MN},
\label{com-2T}
\end{equation} 
we can show that 
\begin{equation}
C_2(Sp(2,\mathbb R) = \frac14\left(X^MP^2X_M-(X\cdot P)(P\cdot X)+\frac{d^2}{4}-1\right).
\label{bas-Cas3}
\end{equation}
On the other hand one defines the quadratic Casimir operator for the $SO(2,d)$ group as 
\begin{equation}
C_2(SO(2,d))=\frac12L_{MN}L^{MN} 
\label{bas-Cas4}
\end{equation}
and the direct calculation shows that 
\begin{equation}
C_2(SO(2,d))=4C_2(Sp(2,\mathbb R)+1-\frac{d^2}{4}.
\label{bas-Cas5}
\end{equation}
Thus, if the generators of the $Sp(2,\mathbb R)$ eliminate quantum states and their quadratic Casimir operator should be equal to zero, the quadratic Casimir operator on the same quantum states treated as belonging to a representation of the $SO(2,d)$ group should be equal to $1-\frac{d^2}{4}$. It is this requirement that fixes the ordering in the generators of the $SO(2,d)$ group  \cite{Bars4}.  

Now, let us note that if we manage to fix the ordering in the generators of $SO(2,d)$ group at the moment when the proper time $\tau$ is equal to zero, then the same ordering will be conserved. At this moment we can note that our parametrization of the variables $X^M,P^M$ at the moment $\tau=0$ coincides with that presented in the paper \cite{Bars4} for the description of the hydrogen atom provided we have already put $E=E_0$.  This fact looks amazing because the physical systems under consideration are quite different and their actions are also different. However, to obtain the action we should consider time-dependent phase variables and the time evolutions  of our two sets of 
variables are quite different. Moreover, introducing the variables as they are presented in the Table 1, we consider $E$ as an independent variable, while $E_0$ is some non-vanishing constant. In this way we manage to obtain the Carroll particle action  \eqref{CarrollActionTrovata}. Then, after writing down the generators of the $SO(2,d)$ group 
\eqref{S:GenRowI}-\eqref{S:GenRowVII}
and their Poisson brackets  we put $E=E_0$ to close the algebra $\mathfrak{so}(2,d)$.

Thus, now we can use the results of the paper \cite{Bars4} to fix the ordering of the operators in the quantum generators of the group $SO(2,d)$:
\begin{align}\label{SQ:GenRowI}
		L^{ij}&=x^{i}p^{j}-x^jp^i,\\\label{SQ:GenRowII} 
		L^{0i}&=\frac{1}{2}\left(r\ p^i+p^i r\right),\\\label{SQ:GenRowIII} 
		L^{+i}&=-E_0 x^i\\\label{SQ:GenRowIV} L^{-i}&=-\frac{1}{2E_0}p_jx^ip^j\cr
		&\qquad\qquad+\frac{1}{2E_0}\left(p\cdot x p^i+p^ix\cdot p\right)+\frac{x^i}{8E_0 r^2} ,\\\label{SQ:GenRowV}
		L^{+-}&=-\frac{1}{2}\left(x\cdot p + p\cdot x\right),\\\label{SQ:GenRowVI}
		L^{-0}&=-\frac{1}{2E_0}p_irp^i-\frac{5-2d}{8E_0r},\\\label{SQ:GenRowVII}
		L^{+0}&=-E_0r.
	\end{align}
The last terms in \eqref{SQ:GenRowIV} and \eqref{SQ:GenRowVI} arise from the ordering choice, which guarantees the closure of the algebra of the generators with respect to the commutators and the disappearance of the quadratic Casimir operator of the group $Sp(2,\mathbb R)$.
   
Now, we shall construct the representation of the $SO(2,d)$ group corresponding to our physical system using the algorithm described in \cite{Bars4} and taking the advantage from the fact that the generators of this group for the Carroll particle and for the hydrogen atom are quite similar. 

First of all, let us define the quadratic Casimir relative to the rotations's subgroup $SO(d-1)$, generated by the $L^{ij}$:
	\begin{align}
		C_2(SO(d-1))&=\frac{1}{2}L_{ij}L^{ij}=\textbf{L}^2\cr 
		&=x_ip^2x^i-(x\cdot p)(p\cdot x).
	\end{align}
	Following \cite{Bars4}, we can observe that the three generators $(L^{+-},L^{+0},L^{-0})$ form an $\mathfrak{so}(1,2)$ subalgebra. Let us define
	\begin{align}
		J^0&=-L^{-0}-\frac{1}{2}L^{+0},\\
		J^1&=-L^{-0}+\frac{1}{2}L^{+0},\\
		J^2&=-L^{+-}.
	\end{align}
	Their commutation rules are given by Eqs. \eqref{basis-Cas1}.
	%\begin{align}
	%	&[J^0,J^1]=iJ^2,\\
	%	&[J^1,J^2]=-iJ^0,\\
	%	&[J^2,J^0]=iJ^1.
	%\end{align}
	The Casimir operator associated with the representation given by our choice of the parameters is
	\begin{align}
		C_2(SO(1,2))&=({J^0})^2-({J^1})^2-({J^2})^2\cr
		&=\textbf{L}^2+\frac{1}{4}(d-2)^2-\frac{d}{2}+1\cr 
		&=j(j+1).
	\end{align}
It is worth observing here that when $d=4$ (namely, when the Carroll system lives in $3+1$ dimensions), the quadratic Casimir of $SO(1,2)$ is the same as the one of the rotations subgroup. Now, let us consider a physical states in which $J^0$ is diagonal. Using the commutation rules \eqref{RQCommRules}, it can be rewritten as
	\begin{align}\
		J^0=r^{\frac{1}{2}}\left(\frac{p^2}{2E_0}+\frac{E_0}{2}\right)r^{\frac{1}{2}}.
	\end{align}
	The eigenstates of this operator are labeled by $j$ and the integer eigenvalues of $J^0$, $m$.
	\begin{align}
		J^0|j,m\rangle = m|j,m\rangle.
	\end{align}
Let us reabsorb a $r^{\frac{1}{2}}$ factor in $|j,m\rangle$ and define $|\Psi_{j,m}\rangle=r^{\frac{1}{2}}|j,m\rangle$. In this way one can observe that
	\begin{align}
		&r^{-\frac{1}{2}}\left(J^0- m\right)|j,m\rangle \cr
		&\qquad=\left(\frac{p^2}{2E_0}-\frac{m}{r}+\frac{E_0}{2}\right)|\Psi_{j,m}\rangle.
	\end{align}
	
	Rescaling by a factor $m^2$ and making a suitable redefinition of the variables $\tilde{r}=mr$ and $\tilde{p}^i=p^i/m$, one obtains 
	\begin{align}
		\left(\frac{\tilde{p}^2}{2E_0}-\frac{1}{\tilde{r}}\right)|\Psi_{j,m}\rangle=\frac{E_0}{2m^2}|\Psi_{j,m}\rangle,
	\label{hydr-like}
	\end{align}
	which is the equation for the hydrogen atom, in which $E_0$ plays the role of the mass.
	
	We may also observe that
	\begin{align}
		r^{-\frac{1}{2}}L^{0+}|j,m\rangle&=H|\Psi_{j,m}\rangle=E_0|\Psi_{j,m}\rangle,\\
		r^{-\frac{1}{2}}L^{0i}|j,m\rangle&=p^i|\Psi_{j,m}\rangle.
	\end{align}
	
	In the paper \cite{Bars4} the similar equation in the case when $d=4$ gives the standard spectrum of the energy eigenvalues of the hydrogen atom. Does it mean that we should have similar discrete spectrum of the energy levels for the Carrol particle? The answer to this question is negative. The point is that the combination of the squared momentum and of the inverse radius which we see in Eq. \eqref{hydr-like} in our case is not connected with the Hamiltonian of the system in contrast to the case of the hydrogen atom. Indeed, our Hamiltonian is simply freely chosen (but different from zero) constant $E_0$, which does not depend on the momentum and the position of the particle. Thus, for us Eq. \eqref{hydr-like}  simply gives some kind of the quantization of this combination. Is it essential? We can say that it does not limit the properties of our quantum Carroll particle at rest, because the momentum $p^i$ is not connected with the velocity of the particle (which is equal to zero). Then, what is the role of the momentum? In quantum theory it enters into the commutation relations \eqref{RQCommRules} and, hence, the standard Heisenberg inequality of uncertainties 
	\begin{equation}
	\Delta x^i\cdots \Delta p^j \geq \frac14\delta^{ij}
	\label{Heis}
	\end{equation}
	is valid. 
	Finally, we can ask ourselves how the Heisenberg relation \eqref{Heis} can be compatible with the fact that the particle should be in rest and should be localized. We believe that here, in contrast to the standard non-relativistic quantum mechanics,
	we can choose the quantum states with a dispersion of the coordinate $\Delta x$ as small as we wish, because the growth of the dispersion of the momentum $\Delta p$ is not important. Thus, a particle can be localized with an arbitrary high precision.

\section{Concluding remarks}

The Carroll group and Carroll symmetry have acquired  a great popularity during last years. Their properties rather interesting from a purely mathematical point of view have found numerous and sometimes unexpected physical applications. The two time physics is less known, but its results are also very interesting because they permit to see quite different physical systems and phenomena from a unified point of view. As far as we know there were no studies devoted to a description of  Carroll particles 
from the point of view of the two time physics. We have made such an attempt in the present paper. Here we have limited ourselves by investigation of a  	relatively simple case of Carroll particles which have non-vanishing energy and should stay at rest in a $(d-1)+1$ spacetime. For this case we have found such  gauge-fixing conditions in the  enlarged $d+2$ dimensional spacetime (which possesses two time variables) which together with the constraints of the theory give a parametrization 
of the phase space variables in the enlarged spacetime that produces Carroll particle in the standard one time spacetime. 

Remarkably, if we treat our phase variables as the quantum operators, then at the moment when the proper time parameter is equal to zero, our parametrization coincides with that obtained in \cite{Bars4} for the hydrogen atom up to some coefficients.
That permits us to follow the quantization scheme developed in \cite{Bars4} for this case. The equations which we obtain are very close to those obtained there, while their physical sense and interpretation are quite different. The roots of this difference lie 
in the fact that our Hamiltonian does not depend neither on the momenta nor on the coordinates of the system. Moreover,
the momenta are not connected with the velocities in contrast with the traditional formulas to which one is accustomed working with Lorentz or Galilei symmetric systems. The role of momenta consists in the fact that they obey the standard commutation   relations 
with the operators of position and, hence, the Heisenberg indeterminacy inequality is valid. However, the role of the dispersions of the coordinates  and of the dispersions of the momenta are different. While the dispersion of the coordinate characterizes the localization of the partilce, the dispersion of the momenta do not have a direct physical sense and one can choose the physical state with an arbitrary high degree of the space localization which is compatible with the fact that classical Carroll particle should always stay in rest. All said above concerns the particles with a non-zero value of energy. The case with zero energy when the Carrol particles are always in motion is more complicated and we hope to present an analysis of this case in the future \cite{future}.

Another possible and interesting direction of the research is to look for  field-theoretical systems with the Carroll symmetry in the two time world.

\begin{acknowledgements}
A.K. is partially supported by the INFN grant FLAG.
\end{acknowledgements}

\end{document}